\documentclass[aps,prl,twocolumn,showpacs,groupedaddress]{revtex4}
\usepackage{graphicx} 
\usepackage{amsmath}
\usepackage{longtable}

\begin{document}

\title{Generating phase shifts from pseudo state energy shifts}
\author{J.Mitroy}
  \email{jxm107@rsphysse.anu.edu.au}
\affiliation{Faculty of Technology, Charles Darwin University, Darwin NT 0909, Australia}
\author{M.W.J.Bromley}
  \email{mbromley@physics.sdsu.edu}
\affiliation{Department of Physics, San Diego State University, San Diego CA 92182, USA}

\date{\today}

\begin{abstract}

A simple way to generate low energy phase shifts for elastic  
scattering using bound-state calculations is postulated, validated and 
applied to the problem of $e^+$-Mg scattering.  The essence of 
the method is to use the energy shift between a small reference 
calculation and the largest possible calculation of the lowest 
energy pseudo-state to tune a semi-empirical optical potential.   
The $L = 1$ partial wave for $e^+$-Mg scattering is predicted to 
have a shape resonance at an energy of about 0.13 eV.  The
value of $Z_{\rm eff}$ at the center of the resonance is 
about 1500.    

\end{abstract}


\maketitle 

One of the most technically demanding problems in quantum physics
is the scattering problem, i.e. the prediction of the reaction 
probabilities when two objects collide \cite{burke95}.  
The underlying difficulty lies in the unbounded nature of 
the wave function.  This leads to a variety of computational and 
analytic complications that are simply absent in bound state 
calculations, e.g. the Schwartz singularities that occur in 
the Kohn variational method for scattering 
\cite{schwartz61a,nesbet80a}.    

One strategy adopted to solve scattering problems is to use bound state 
methods.  There are numerous examples of such approaches, one of the 
most popular being the $R$-matrix methods
that rely on the solutions of the Schrodinger equation in a
finite sized cavity to determine the behaviour of the wave function
in the interaction region \cite{burke95}.  The total wave function is 
then constructed by splicing the inner wave function onto the asymptotic 
wave function.           

However, despite the considerable activity in this area, there 
are a number of problems that are beyond resolution.  
The $e^+$-atom problem is a notoriously hard numerical problem since 
the atomic electrons tend to localize around the positron, thus 
giving a very slowly convergent partial wave expansion of the wave 
function inside the interaction region (this should not be confused 
with the partial wave expansion of the asymptotic wave function) 
\cite{higgins90,mitroy99c,mitroy06a,armour06c}.   
For example, the dimensionality of the equations to be solved
to achieve a given accuracy are about 5 times larger for 
$e^+$-H scattering than for $e^-$-H scattering.  At present,
there are a number of positron collision problems that are
simply inaccessible with existing approaches \cite{armour06c}.  

This article had its origin in a particular scattering problem, 
namely the determination of the near threshold phase shifts for 
positron scattering from the di-valent group II and IIB atoms. The
dimension of the secular equations for bound state calculation 
on such systems are very large, for example a CI calculation of 
the $e^+$Ca $^2$P$^{\rm o}$ state resulted in equations of
dimension 874,448 \cite{bromley06f}.  Application of the CI-Kohn
approach \cite{bromley03a} to determine the phase shifts for
$e^+$-Mg scattering in the $^2$P$^{\rm o}$ channel would result in 
linear equations that are simply too large ($\approx 1,000,000$)  
to be solved by direct methods.  Iterative methods do exist,  but 
there are no robust methods that absolutely guarantee convergence  
\cite{saad00a}. It is likely that the development of an efficient 
linear solver for the class of problems that arise from a basis 
set treatment of quantum scattering would involve a good deal 
of initial effort and experimentation.  There is, however,  
a great deal of experience in obtaining the lowest eigenvalues 
of large symmetric matrices \cite{stathopolous94a}.  
   
The idea behind the current method lies closest to the trivial 
$R$-matrix method \cite{percival57a} which is exploited in Quantum 
Monte Carlo (QMC) calculations of scattering \cite{alhassid84a}. In 
the QMC, one extracts the phase shift by comparing the zero point
energy of a finite size cavity to the energy of the system wave 
function in the same cavity.  In the present method, the phase shift 
is extracted from the energy shift when a reference wave function is 
enlarged in size to account for short and long range correlations.    
The method is applied to $e^+$-Mg scattering in the $^2$P$^{\rm o}$ 
symmetry and used to predict the existence of a prominent shape 
resonance at 0.13 eV incident energy.  This is noteworthy since 
shape resonances are currently unknown in $e^+$-atom or 
$e^+$-molecule scattering \cite{surko05a}.  

Our method proceeds as follows. The initial calculation uses a 
reference CI wave function of  product form, viz 
\begin{equation}
\Psi_0  = \Phi_{\rm gs}({\mathbf X})  \phi_{0}({\mathbf r}) \ . 
\label{refwf}
\end{equation}
The wave function of the parent atom is 
$\Phi_{\rm gs}({\mathbf X})$ where ${\mathbf X}$ is the 
collective set of target coordinates.  The wave function 
of the projectile is $\phi_{0}({\mathbf r})$. 
In general, $\phi_{0}$ is a linear combination of a finite 
number of square-integrable functions designed to give a good 
representation of the wave function in a bounded interaction 
region.  The energy expectation, $E_0$ is given by 
\begin{equation}
E_0 = \langle \Psi_0 |  H_{\rm exact} | \Psi_0 \rangle .   
\label{refE}
\end{equation}

The wave function $\Psi_0$ is then augmented by a very 
large number of additional functions to represent the
correlations between the projectile and the
target constituents.  This augmented trial function
is 
\begin{equation}
\Psi_1  = \Phi_{\rm gs}({\mathbf X}) \phi_{0}({\mathbf r})  
+ \sum_{i,j} c_{i,j} \Phi_{\rm i}({\mathbf X})  \phi_{j}({\mathbf r}) \ . 
\label{exactwf}
\end{equation}
The trial wave function $\Psi_1$ is used to diagonalize $H_{\rm exact}$
giving an energy of $E_1$.  The additional functions do not include   
any that have the same sub-symmetries as those comprising $\Psi_0$.  

\begin{figure}[tbh]
\centering{
\includegraphics[width=8.0cm,angle=0]{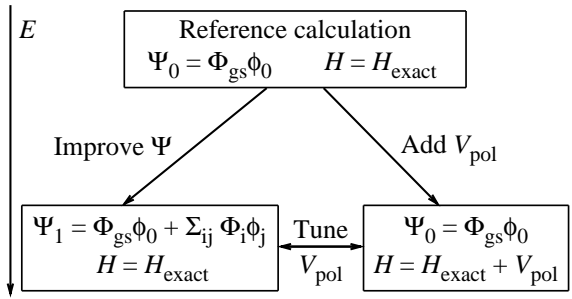}
}
\caption[]{ \label{Schematic}
Schematic diagram showing the strategy used to  
tune the semi-empirical optical potential.  
}
\end{figure}

Next, a semi-empirical potential of the form  
\begin{equation}
V_{\rm pol}  = \frac{\alpha_d}{2r^4} \left( 1 - \exp(-r^6/\rho^6) \right) \ ,  
\label{vpol}
\end{equation}
is added to $H_{\rm exact}$ ($\alpha_d$ is the dipole polarizability).  
This potential only acts on the scattering projectile.  Then $\Psi_0$ 
is used to diagonalize $H_{\rm exact} + V_{\rm pol}$ 
giving $E_{\rm pol}$.  The parameter $\rho$ in eq.~(\ref{vpol})
is adjusted until $E_{pol} = E_1$.  Figure \ref{Schematic} is a schematic 
diagram outlining this procedure. 

In the final stage, $\Psi_0$ is modified to permit  
$\phi_{0}({\mathbf r})$ to describe continuum solutions,    
\begin{equation}
\Psi_{\rm continuum}  = \Phi_{\rm gs}({\mathbf X})  \phi_{\rm continuum}({\mathbf r}) \ .  
\label{refwf2}
\end{equation}
The phase shifts of $H_{\rm exact} + V_{\rm pol}$
are then obtained by using $\Psi_{\rm continuum}$ as the
scattering wave function.

The method is verified by computing the low energy phase
shifts and annihilation parameters for $s$-wave $e^+$-H 
scattering.  The reference wave function, $\Psi_0$,  consisted of 
the hydrogen atom ground state multiplied by a positron basis 
of 30 $\ell = 0$ Laguerre type orbitals.  The energy and 
annihilation rate of $\Psi_0$ are given in Table \ref{hlmax}.    
  
\begin{table}[th]
\caption[]{  \label{hlmax}
Results of CI calculations for the $^1$S$^{\rm e}$ symmetry 
of $e^+$H for a series of $J$.  The number of electron ($N_e$)
and positron ($N_p$) orbitals are listed.  The total number of two-body 
functions in the CI basis are in the $N_{CI}$ column.  
Energies are given in Hartree while spin-averaged annihilation 
rate ($\Gamma$) are given in units of $10^9$ s$^{-1}$ ($\Gamma$ 
for $\Psi_0$ 
is for the tuned $V_{pol}$). Also given are the extrapolations
to the $J \to \infty$ limits using eq.~(\ref{extrap1}).
}
\begin{ruledtabular}
\begin{tabular}{lccccc} 
$J$    &  $N_e$ & $N_p$ & $N_{CI}$ & $\langle E \rangle_J$ & $\langle \Gamma \rangle_J$  \\
\hline
$\Psi_0$ & 1 &  30 &    30 & -0.49772560 & 0.00089605   \\
    9  & 250 & 259 & 6511  & -0.49797210 & 0.0040914253 \\  
   10  & 274 & 283 & 7087  & -0.49797276 & 0.0042047713 \\
   11  & 298 & 307 & 7663  & -0.49797325 & 0.0042994659 \\  
   12  & 322 & 331 & 8239  & -0.49797360 & 0.0043795165 \\  
\multicolumn{6}{c}{$J \to \infty$ extrapolations}   \\  \hline  
\multicolumn{4}{l}{1-term eq.~(\ref{extrap1})}  & -0.49797439 &  0.005341190   \\  
\multicolumn{4}{l}{2-term eq.~(\ref{extrap1})}  & -0.49797509 &  0.005334089   \\ 
\multicolumn{4}{l}{3-term eq.~(\ref{extrap1})}  & -0.49797509 &  0.005264739   \\  
\end{tabular}
\end{ruledtabular}
\end{table}

A sequence of successively larger calculations with $J$ (the maximum
$\ell$ value of any orbital included in the basis) were done up
to $J = 12$.  The energies at a given $J$, $\langle E \rangle_J$,  
and annihilation rates, $\langle \Gamma \rangle_J$, are given in
Table \ref{hlmax}.  A major problem affecting CI calculations of 
positron-atom interactions is the slow convergence of the energy 
with $J$ \cite{mitroy99c,mitroy02b,mitroy06a}.  
One way to determine the $J \rightarrow \infty$ energy, $\langle E \rangle_{\infty}$,
is to make use of an asymptotic analysis.  It has been shown that successive 
increments, $\Delta E_{J} = \langle E \rangle_J - \langle E \rangle_{J-1}$, 
to the energy can written as an inverse power series 
\cite{schwartz62a,carroll79a,hill85a,mitroy06a,bromley06a}, viz 
\begin{equation}
\Delta E_L \approx \frac {A_E}{(L+{\scriptstyle \frac{1}{2}})^4} 
    + \frac {B_E}{(L+{\scriptstyle \frac{1}{2}})^5} 
    + \frac {C_E}{(L+{\scriptstyle \frac{1}{2}})^6} + \dots \ \   .
\label{extrap1}
\end{equation}
The $J \to \infty$ limits have been determined by fitting sets of 
$\langle E \rangle_J$ values to asymptotic series with either 1, 2 
or 3 terms.  The linear factors, $A_E$, $B_E$ and $C_E$ for the 
3-term expansion are determined at a particular $J$ from 4 successive 
energies ($\langle E \rangle_{J-3}$, $\langle E\rangle_{J-2}$, 
$\langle E \rangle_{J-1}$ and $\langle E \rangle_{J}$).  
Once the linear factors have been determined it is trivial to sum 
the series to $\infty$ \cite{salomonson89b,mitroy06a,bromley06a}
(the $J \to \infty$ limits are given in Table \ref{hlmax}).  

\begin{figure}[tbh]
\centering{
\includegraphics[width=8.6cm,angle=0]{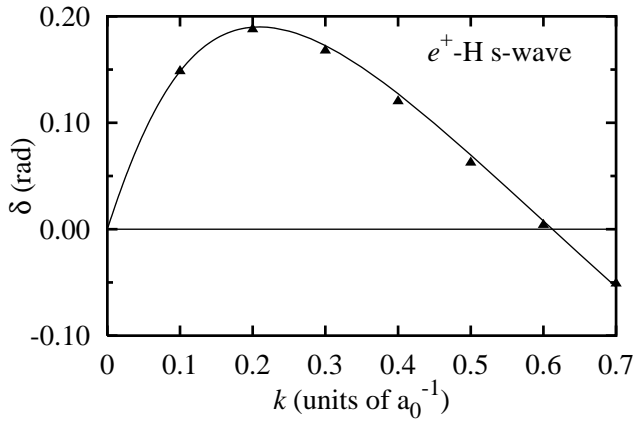}
}
\caption[]{ \label{Hphase}
The phase shift for $e^+$-H scattering in the $s$-wave as a 
function of $k$ (in units of $a_0^{-1}$).  The solid line
shows the results of the present calculation while the 
triangles show the close to exact phase shifts of
Bhatia {\em et al} \cite{bhatia71}.
}
\end{figure}

The trial function $\Psi_0$ was then used to diagonalize the 
Hamiltonian with an additional polarization potential 
($\alpha_d = 4.5$ $a_0^3$).  The energy from this calculation
matches the 3-term extrapolation in Table \ref{hlmax} when
$\rho = 2.0495 \ a_0$.  This value of $\rho$ is very close
to a value of $\rho = 2.051 \ a_0$ that was obtained when a 
polarization potential of this form was tuned to a very accurate 
phase shift in a semi-empirical investigation of $e^+$-H 
scattering \cite{mitroy02a}.  The phase shifts obtained 
by integrating the Schrodinger equation for the model Hamiltonian 
with $\rho = 2.0495 \ a_0$ are depicted in Figure \ref{Hphase}
and the level of agreement with the close to exact phase shifts
could hardly be better. 
 
Besides obtaining phase shifts, this procedure was used to
determine the annihilation parameter, $Z_{\rm eff}$.  In this case the
extrapolation to the $J \to \infty$ limits were done with 
an asymptotic series similar as eq.~(\ref{extrap1}) but with 
the leading order starting as $A_{\Gamma}/(J+1/2)^2$.  The 
ratio between the annihilation rates calculated with $\Psi_0$ 
and $\Psi_1$ can be equated with the enhancement factor, $G$, 
for $s$-wave $e^+$-H scattering \cite{mitroy02a}. 
The 2-term extrapolation is chosen for the evaluation of the
ratio since lack of completeness in the finite dimension radial 
basis will have a bigger effect on the 3-term extrapolation
\cite{bromley06a,mitroy06a}.  With this choice the enhancement
factor becomes $G = 5.95$, which is within 1.5$\%$ of the enhancement 
factor chosen by normalization to a very accurate $T$-matrix 
close coupling calculation \cite{ryzhikh00a,mitroy02a}.  
The predicted $Z_{\rm eff}$, although not shown, lie within
5$\%$ of those of Bhatia {\em et al} \cite{bhatia74b} over the 
$k \in [0,0.7] \ a_0^{-1}$ range.   

This approach to computing the phase shifts was applied to the 
determination of $e^+$-Mg scattering in the $L_T = 1$ partial 
wave.  The treatment of Mg requires the use of a frozen core 
approximation whose details have been discussed  
elsewhere \cite{bromley02b,mitroy06a}, so only the 
briefest description is given here.  The model Hamiltonian 
is initially based on a Hartree-Fock (HF) wave function for 
the Mg ground state.  The core orbitals are then 
frozen.  The impact of the direct and exchange part of the HF 
core interactions on the active particles are computed without 
approximation.  One- and two-body semi-empirical core-polarization 
potentials are then added to the potential.  The adjustable 
parameters of the core-polarization potential are defined by 
reference to the spectrum of Mg$^+$ \cite{bromley02b}.

The $e^+$Mg CI basis was constructed by letting the two electrons and 
the positron form all the possible configuration with a total angular 
momentum of $L_T = 1$, with the two electrons in a spin-singlet 
state, subject to the selection rules,
\begin{eqnarray}
\max(\ell_0,\ell_1,\ell_2) & \le & J \ , \\
\min(\ell_1,\ell_2)& \le & L_{\rm int} \ ,  \\  
(-1)^{(\ell_0+\ell_1+\ell_2)}& = & -1 \ . 
\end{eqnarray}
In these rules $\ell_0$, $\ell_1$ and $\ell_2$ are respectively 
the orbital angular momenta of the positron and the two electrons.  

\begin{table}[th]
\caption[]{  \label{tab:mglmax}
Results of CI calculations for the $^2$P$^{\rm o}$ state of $e^+$Mg 
for a series of $J$ ($L_{\rm int}=3$).   The threshold 
for binding is -0.83285190 Hartree.  Most aspects of the Table 
are similar to those of Table \ref{hlmax}.
}
\begin{ruledtabular}
\begin{tabular}{lccccc} 
$L_{\rm max}$&  $N_e$ & $N_p$ & $N_{CI}$ & $\langle E \rangle_J$ & $\langle \Gamma \rangle_J$    \\
\hline
$\Psi_0$  &   & 20  & 20  & -0.82525710 &  0.029828  \\  
11 & 172 & 174 & 651006 & -0.82806307 & 0.12800208 \\  
12 & 186 & 188 & 724506 & -0.82817969 & 0.14306354 \\  
13 & 200 & 202 & 798006 & -0.82827695 & 0.15662562 \\  
14 & 214 & 216 & 871506 & -0.82835799 & 0.16873961 \\ \hline  
\multicolumn{6}{c}{$J \to \infty$ extrapolations}  \\  
\multicolumn{4}{l}{1-term eq.~(\ref{extrap1})}  & -0.82871101 &  0.338475  \\  
\multicolumn{4}{l}{2-term eq.~(\ref{extrap1})}  & -0.82884022 &  0.373490  \\ 
\multicolumn{4}{l}{3-term eq.~(\ref{extrap1})}  & -0.82886332 &  0.315877  \\  
\end{tabular}
\end{ruledtabular}
\end{table}

The Hamiltonian for the $e^+$Mg $^2$P$^{\rm o}$ state was
diagonalized in a CI basis constructed from a large number of 
single particle orbitals, including orbitals up to $\ell = 14$.
The two electrons were in a spin singlet state.  There was a 
minimum of $14$ radial basis functions for each $\ell$.  There 
were 20 $\ell = 1$ positron orbitals.  The largest calculation 
was performed with $J = 14$ and $L_{\rm int} = 3$.  The parameter 
$L_{\rm int}$ was set to $L_{\rm int} = 3$ since this is mainly 
concerned with describing the more quickly converging electron-electron 
correlations \cite{bromley02b}.  The secular equations were 
solved with the Davidson algorithm \cite{stathopolous94a}.

\begin{figure}[tbh]
\centering{
\includegraphics[width=9.0cm,angle=0]{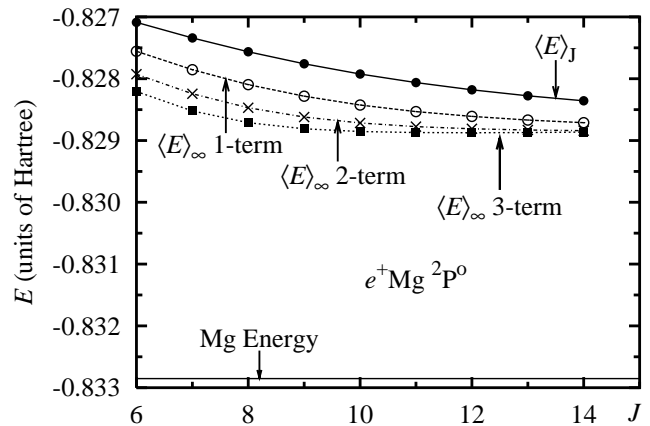}
}
\caption[]{ \label{fig:MgE}
The energy of the $^2$P$^{\rm o}$ state 
of $e^+$Mg as a function of $J$.  The directly calculated 
energy is shown as the solid line while the $J \to \infty$ 
limits using eq.~(\ref{extrap1}) with 1, 2 or 3 terms are shown 
as the dashed lines.  The Mg + $e^+$ dissociation threshold is 
shown as the horizontal line.
}
\end{figure}

First, it is necessary to get the Mg ground state energy in 
this basis.  The limitation $L_{\rm int} = 3$ means that only 
a single electron in the model atom can have $\ell > 3$.  
Translating this to an equivalent CI calculation for the Mg ground
state resulted in an energy of of $E = -0.83285190$ Hartree 
(energy given relative to the Mg$^{2+}$ core).

The energy and annihilation rate of the $e^+$Mg 
$^2$P$^{\rm o}$ state as a function of $J$ are given in 
Table \ref{tab:mglmax}.  Figure \ref{fig:MgE} shows the running 
estimates of $\langle E \rangle_{\infty}$ with the $J \to \infty$ 
extrapolations as a function of $J$.   It is clear that none of 
calculations indicate the existence of a bound state, but the 
energy shift algorithm has to be applied to determine whether 
this is due to the finite basis size.  

\begin{figure}[tbh]
\centering{
\includegraphics[width=8.7cm,angle=0]{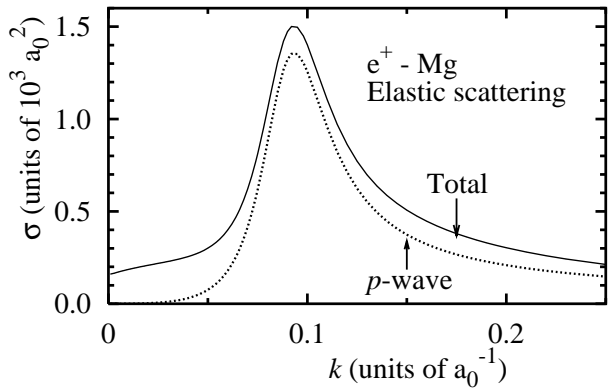}
}
\caption[]{ \label{Mgcross}
The elastic scattering cross section for $e^+$-Mg scattering 
in energy region below the Ps-formation threshold.
The solid line shows the total cross section while the
dashed curve shows the partial cross section of the $\ell = 1$ 
partial wave.
}
\end{figure}

A polarization potential given by eq.~(\ref{vpol}) with 
$\alpha_d = 72 \ a_0^3$ \cite{mitroy02a,bromley02b} 
(the Mg ground state polarizability) was added to original 
Hamiltonian and $\rho$ was tuned until an energy shift of 
0.003588 $(= -0.82886332 + 0.82525710)$ Hartree was achieved. Figure 
\ref{Mgcross} shows the elastic cross section for $e^+$-Mg scattering 
below the Ps formation threshold (at $k \approx 0.25 \ a_0^{-1}$).   
The cutoff parameters in eq.~(\ref{vpol}) were set to 
$\rho = 3.032 \ a_0$ for the $s$-wave \cite{mitroy02a} and 
$\rho = 2.573 \ a_0$ (derived here) for all the other partial 
waves.  The elastic cross section in this energy region 
is almost completely dominated by a $p$-wave 
shape resonance with its center near $k \approx 0.10$ $a_0^{-1}$. 

The value of $Z_{\rm eff}$ at the resonance peak was about 
1500.  This was determined by using an enhancement factor 
of $G = 12.5 = 0.3735/0.02983$ for valence annihilation.
It is likely that this is an underestimate since a 
lack of completeness in the finite dimension radial 
basis usually leads to annihilation rates being too
small \cite{mitroy06a}. 

To summarize, a novel technique has been used to demonstrate the 
existence of a shape resonance in $e^+$Mg scattering which has 
the virtue of being readily detectable.  The phase shift calculations 
were performed using a semi-empirical method \cite{mitroy02a} with 
a tuned potential.  The tuning of a semi-empirical optical potential 
to features such as bound state energies and resonance positions 
is well known.  The unique feature of the present approach is that 
the optical potential is tuned to the energy shift of a positive energy 
pseudo-state.  This approach to the calculation of phase 
shifts can be applied to other scattering systems which
are currently inaccessible using existing techniques.

The calculations upon the $e^+$Mg system were performed on Linux
clusters hosted at the SDSU Computational Sciences Research Center 
and the South Australian Partnership for Advanced
Computing (SAPAC) and the authors would like to thanks to Grant
Ward and Dr. James Otto for their assistance. The authors also 
thank Prof. Bob McEachran for a critical reading of the manuscript. 


\end{document}